\documentclass[graybox]{svmult}

\usepackage{type1cm}        
\usepackage{makeidx}         
\usepackage{graphicx}        
\usepackage{multicol}        
\usepackage[bottom]{footmisc}
\usepackage[round]{natbib}

\usepackage{lipsum}
\usepackage{newtxtext}       %
\usepackage{newtxmath}       

\makeindex             

\begin{document}

\title*{An Assessment of the AI Regulation Proposed by the European Commission}
\author{Patrick Glauner}
\institute{Patrick Glauner \at Deggendorf Institute of Technology, Deggendorf, Germany \email{patrick@glauner.info} \\~\\~ Preprint. To appear in the 2022 Springer book ``\textit{The Future Circle of Healthcare: AI, 3D Printing, Longevity, Ethics, and Uncertainty Mitigation}" edited by Sepehr Ehsani, Patrick Glauner, Philipp Plugmann and Florian M. Thieringer.}
%
%
\maketitle

\abstract{In April 2021, the European Commission published a proposed regulation on AI.  It intends to create a uniform legal framework for AI within the European Union (EU). In this chapter, we analyze and assess the proposal. We show that the proposed regulation is actually not needed due to existing regulations. We also argue that the proposal clearly poses the risk of overregulation. As a consequence, this would make the use or development of AI applications in safety-critical application areas, such as in healthcare, almost impossible in the EU. This would also likely further strengthen Chinese and US corporations in their technology leadership. Our assessment is based on the oral evidence we gave in May 2021 to the joint session  of the European Union affairs committees of the German federal parliament and the French National Assembly.}

\section{Introduction}
\begin{figure}[h!]
\centering
\includegraphics[width=\textwidth]{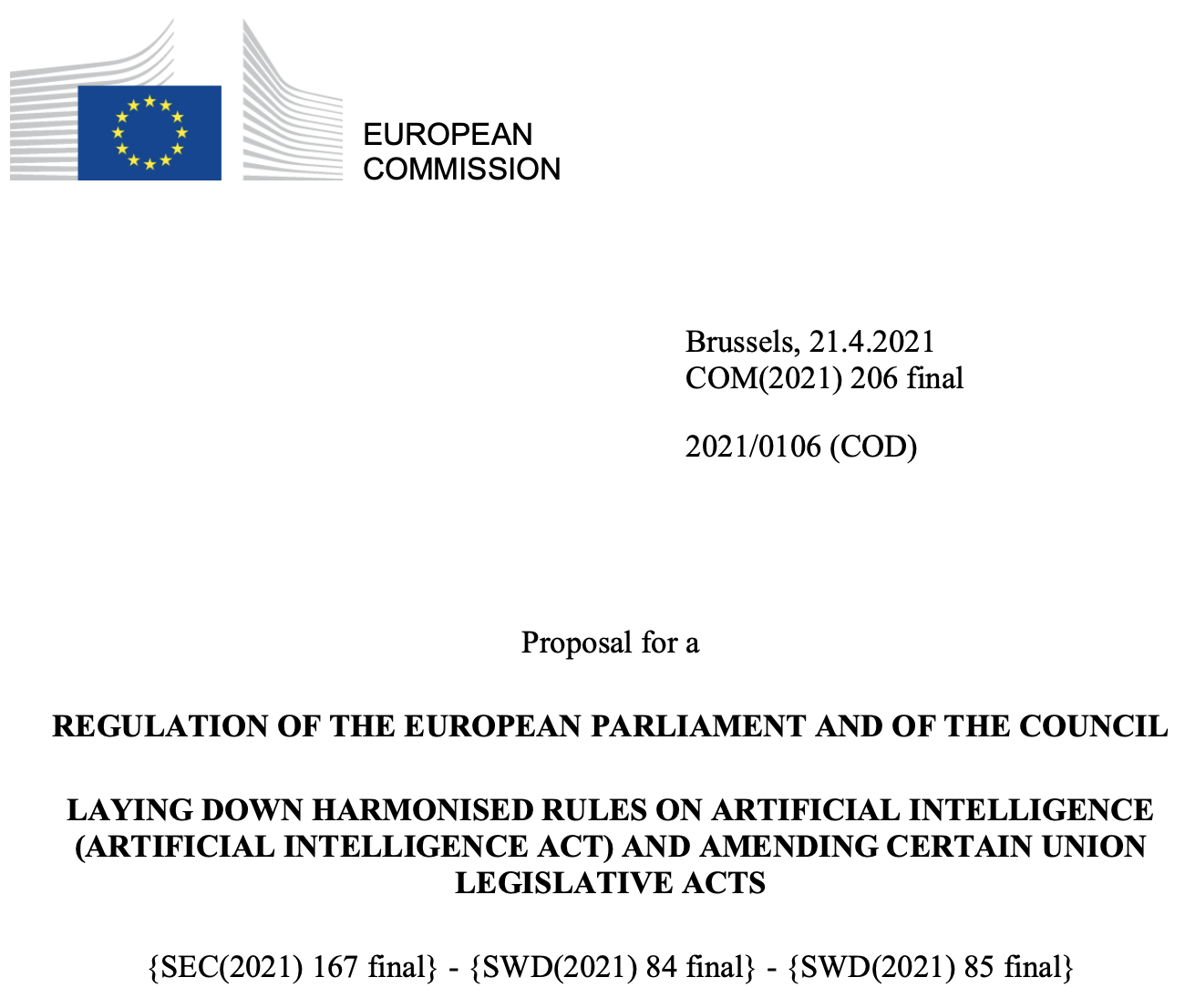} 
\caption{EU Proposal for a Regulation Laying Down Harmonised Rules on Artificial Intelligence (Artificial Intelligence Act). Source: \citep{eu2021proposal}.}
\label{fig:euregulation}
\end{figure}

Artificial intelligence (AI) aims to automate human decision-making behavior and is therefore also considered the next phase of the industrial revolution. We have previously reviewed the state of the art and challenges of AI applications in healthcare \citep{glauner2021artificial}. This book provides a forecast of how AI and other technologies are likelty to skyrocket healthcare in the foreseeable future.
The European Commission published proposed regulation in April 2021 \citep{eu2021proposal} that intends to create a uniform legal framework for AI within the European Union (EU). The proposal particularly addresses safety-critical applications, a category that also includes most healthcare applications of AI. The proposal's cover page is depicted in Fig.~\ref{fig:euregulation}.

Prior to that, the European Commission published an AI white paper in February 2020 \citep{eu2020whitepaper}. The white paper intends to serve as an initial step towards an AI strategy for the EU. However, the white paper lacks crucial points: For example, it does not contrast its proposed actions to those of the contemporary AI leaders. The term ``China" does not appear in the white paper at all. Furthermore, the proposed investments appear to be negligible compared to the Chinese investment.

In this chapter, we provide an assessment\footnote{Our professional background: At Deggendorf Institute of Technology, we teach and conduct research on AI. In addition, we advise companies on the use of AI. That activity includes the definition and implementation of AI strategies, the implementation of AI applications, their operation as well as running training courses.} of the EU Commission's AI regulation proposal. Our assessment is based on the oral evidence we gave to the joint session\footnote{See video: \url{http://videos.assemblee-nationale.fr/video.10738054\_6093a70d0eb8d.commission-des-affaires-europeennes--reunion-commune-avec-la-commission-des-affaires-europeennes-du-6-mai-2021} (in French).} of the European Union affairs committees of the German federal parliament and the French National Assembly on May 6, 2021. We also submitted written evidence to the committees \citep{glauner2021stellungnahme}.
Concretely, we show that the proposed regulation poses the risk of overregulation. This would make the use or development of AI applications in safety-critical application areas, such as in healthcare, in the EU almost impossible. We furthermore provide concrete recommendations that the EU should implement in order to foster a sustainable ecosystem that allows European companies to thrive.

\section{Analysis and Assessment}
In this section, we provide an analysis of the proposal and assess its most critical parts. We also argue that a number of critical points have not been addressed at all in the proposal.

\subsection{Too Broad Definition of AI}
There is currently no exact and generally accepted definition of AI, neither in science nor in industry, because of the intersections with other fields, such as statistics and signal processing\footnote{Many popular AI methods also fall into the field of signal processing. However, there have been no correspondingly broad use case-independent regulation attempts in the history of signal processing.}. This shortcoming fundamentally complicates any kind of AI regulation.

The proposed regulation contains its own definition of AI in \textit{Article 3 (Definitions)}:
\newline
\newline
\textit{[...] \\~
(1) ‘artificial intelligence system’ (AI system) means software that is developed with one or more of the techniques and approaches listed in Annex I and can, for a given set of human-defined objectives, generate outputs such as content, predictions, recommendations, or decisions influencing the environments they interact with; \\~
[...]},
\newline
\newline
with reference to \textit{Annex I}:
\newline
\newline
\textit{(a) Machine learning approaches, including supervised, unsupervised and reinforcement
learning, using a wide variety of methods including deep learning; \\~
(b) Logic- and knowledge-based approaches, including knowledge representation,
inductive (logic) programming, knowledge bases, inference and deductive engines,
(symbolic) reasoning and expert systems; \\~
(c) Statistical approaches, Bayesian estimation, search and optimization methods.}
\newline
\newline
In summary, it includes any software\footnote{Note that AI can in principle also be implemented through purely hardware-based approaches without software. The proposed regulation ignores this and would not regulate such AI implementations at all.} that uses machine learning methods or logic-based procedures. However, it also includes any software that uses statistical methods or search and optimization methods. This classifies almost all existing and future software that is not actually AI-related as ``AI".

\begin{tips}{Example}
Any software that computes the mean of numbers uses a ``statistical procedure". As a consequence, the proposal classifies it as ``AI".
\end{tips}

Since almost all software is covered by that set of rules, unforeseeable risks would arise for every company as soon as it uses or develops any kind of software.

\subsection{No Need for AI-Specific Regulation}
Existing regulations, laws, standards, norms, etc. of technologies are in most cases vertically structured. They generally consider systems that address certain safety-critical use cases, such as in  healthcare, aviation or nuclear power. Also, they typically pay little to no attention to the concrete implementation of these systems through hardware, software or a combination thereof.

AI applications are usually only a small part of larger software/hardware systems. The proposed regulation attempts to regulate this proportion horizontally - and thus independently of use cases. This approach appears to be impractical due to the overall mostly uncritical AI use cases. It can be assumed that additional horizontal regulation would also lead to unclear responsibilities and disputes over responsibilities. Additional regulations should therefore only address novel use cases that are not yet covered by existing regulations. 
 
The German AI Association is in favor of applying existing use case-specific regulations and adapting them to AI-specific requirements if necessary \citep{kib2021position}. Bitkom, the federation of German digital companies, is also of the opinion that even a more general horizontal regulation of ``algorithmic systems" is not practicable \citep{bitkom2021stellungnahme}.

\subsection{Lack of Delimitation from Existing Regulations}
The application areas prohibited in \textit{Article 5} of the proposed regulation are very broadly defined and lead to uncertainties for all stakeholders due to the correspondingly wide scope for interpretation\footnote{For example, even operating a search engine on a potentially unrepresentative database could be prohibited by \textit{Article 5(a)} due to the potential impact of the search results.}. Instead, the definition of specifically prohibited use cases would be much more effective. It should also be examined whether these would have to be explicitly prohibited at all or whether this is already the case today under other laws, such as criminal codes or the General Data Protection Regulation (GDPR) \citep{eu2016data}. New prohibitions should also address the relevant use cases in general, and without any reference to AI, as they could (in the future) possibly be implemented without the explicit use of AI methods.

For the same reasons, the definition of concrete use cases without explicit reference to AI would therefore also be helpful in \textit{Article 6 (Classification rules for high-risk AI systems)} for a more precise definition of ``high-risk applications" \citep{politico2021AI}. It is also unclear from the proposed regulation what the European Commission specifically means by ``safe" applications. Furthermore, it is not defined how ``safety" could be established in individual use cases at all, especially since many machine learning methods have statistical uncertainty embedded in their definition.

In conjunction with existing regulations, the proposed regulation also poses the risk of contradictory and twofold requirements, for example in automated lending \citep{handelsblatt2021Neu}.

\subsection{Unfulfillable Requirements for ``High-Risk Applications"}

In addition to the very broad definition of ``high-risk applications"\footnote{I would like to thank in particular Tobias Manthey of EvoTegra GmbH for our extensive discussions on the topic of ``high-risk applications".} in \textit{Article 6}, the proposed regulation provides for corresponding documentation requirements in \textit{Article 11 (Technical documentation)}, registration requirements in \textit{Article 60 (EU database for stand-alone high-risk AI systems)} and reporting requirements in \textit{Article 62 (Reporting of serious incidents and of malfunctioning)}. These requirements for the development or use of AI in safety-critical application areas are comparable to the operation of nuclear power plants or the development of aircraft. They therefore will likely inhibit innovation and are thus disproportionate.

To implement the procedures described in \textit{Article 64 (Access to data and documentation)}, the European Commission would practically have to bundle the EU-wide AI expertise of all companies, universities and experts in the relevant authorities and invest hundreds of billions of euros in its own infrastructure. The requirements for the sandbox tests described in \textit{Article 53 (AI regulatory sandboxes)} state that the entire intellectual property consisting of data and AI would have to be shared with the relevant authorities. This seems disproportionate and unfeasible. In addition, there are open questions about liability should third parties gain access to the intellectual property through sandbox testing.

For these reasons, the proposed regulation would make the development or deployment of relevant safety-critical AI applications, such as any safety-critical assistance systems in healthcare, nearly impossible in the European Union. 

\subsection{Overregulation would Strengthen Chinese and US Corporations}
One of the goals of the GDPR passed in 2016 was to limit the power of US cloud providers and strengthen European companies. In the meantime, however, it has become clear that exactly the opposite has occurred \citep{bloomberg2018facebook}. In particular, the large US cloud providers have the human and financial resources to implement GDPR-compliant services. They also have the financial resources to settle any fines or avoid them by entering complex and lengthy trials.

The proposed regulation would therefore result in European companies being unable to reach or hold a leadership position in international competition in the future due to overregulation. Chinese and US providers of AI-based services and products would be strengthened by this regulation. As a result, Chinese corporations in particular could squeeze European companies out of the market or take them over in every sector in the medium term \citep{lee2018ai}.

\subsection{Missing Points}
While the proposal is very broad, a number of critical points, including but not limited to the following, have not been (adequately) addressed in it:
\begin{itemize}
    \item Liability
    \item Grandfathering and transitional arrangements
    \item Military applications
\end{itemize}

We would also like to highlight that the proposed regulation has been significantly toned down compared to the previous drafts with regard to the use of AI in social networks. It no longer specifically addresses social networks, even though this is precisely where there is a realistic danger of AI influencing our society \citep{cnbc2018facebook}.

These points need to be discussed in greater detail by the European Commission when revising the proposal.

\section{Recommendations}
The European Commission should focus its work on the added value of AI for citizens - especially from the perspective of the expected further increase in prosperity and quality of life. In doing so, it should refrain from overregulating AI.

Broad AI qualification measures should be created across Europe through a European Hightech Agenda. This would make citizens much more shapers of the digital transformation than being shaped by it. In addition, transfer projects between universities and industry partners should be funded in a more targeted and effective manner. Also, a top European AI research institution should be established along the lines of the European Organization for Nuclear Research (CERN) \citep{sciencebusiness2021call}.

The Bavarian Hightech Agenda\footnote{\url{http://www.bayern.de/politik/hightech-agenda/}} should serve as a positive role model. Through this, a variety of AI measures are currently being implemented to strengthen Bavaria-wide competitiveness, such as:
\begin{itemize}
\item AI courses and degree programs\footnote{\url{http://idw-online.de/de/news765811}}.
\item Societal projects, such as DeinHaus 4.0\footnote{\url{http://deinhaus4-0.de/}} for longer and healthier living at home.
\item Transfer projects between universities and startups, such as between Deggendorf Institute of Technology and EVOMECS GmbH for AI-based production planning\footnote{\url{http://www.evomecs.com/wp-content/uploads/2020/10/131020-Presse\_Final\_GER.pdf}}.
\item AI transfer centers in rural areas, including the AI center in the Denkwelt Oberpfalz of the LUCE Foundation\footnote{\url{http://www.luce-stiftung.de/kooperationsvertrag-zwischen-oth-amberg-weiden-und-der-luce-stiftung/}}.
\end{itemize}

\section{Conclusions}
The proposed regulation published by the European Commission in April 2021 intends to create a uniform legal framework for artificial intelligence (AI) within the European Union (EU). The very broad definition of ``AI" contained therein classifies almost any existing and future software as ``AI" and would then be covered by this regulatory framework. However, there is no need for AI-specific regulation due to existing regulations - apart from possibly a few novel use cases. The proposed regulation poses the risk of overregulation, which would make the use or development of AI applications in safety-critical application areas, such as in healthcare, almost impossible in the EU. This would also likely further strengthen Chinese and US corporations in their technology leadership.


\bibliographystyle{apalike}
\bibliography{refs}

\begin{thebibliography}{}

\bibitem[Bershidsky, 2021]{bloomberg2018facebook}
Bershidsky, L. (2021).
\newblock {E}urope’s {P}rivacy {R}ules are {H}aving {U}nintended
  {C}onsequences.
\newblock
  http://www.bloomberg.com/opinion/articles/2018-11-14/facebook-and-google-aren-t-hurt-by-gdpr-but-smaller-firms-are.
\newblock [Online; accessed May 17, 2021].

\bibitem[Bitkom, 2020]{bitkom2021stellungnahme}
Bitkom (2020).
\newblock {S}tellungnahme zum {A}bschlussbericht der {DEK}: {K}urzfassung
  {B}itkom {S}tellungnahme zum {G}utachten und den {E}mpfehlungen der
  {D}atenethikkommission.
\newblock
  http://www.bitkom.org/sites/default/files/2020-04/20200402\_kurzfassung-bitkom-stellungnahme-zum-abschlussbericht-der-dek.pdf.
\newblock [Online; accessed May 17, 2021].

\bibitem[{European Commission}, 2016]{eu2016data}
{European Commission} (2016).
\newblock {T}he {G}eneral {D}ata {P}rotection {R}egulation ({GDPR}).
\newblock
  http://ec.europa.eu/info/law/law-topic/data-protection/data-protection-eu\_en.
\newblock [Online; accessed May 24, 2021].

\bibitem[{European Commission}, 2020]{eu2020whitepaper}
{European Commission} (2020).
\newblock {W}hite {P}aper on {A}rtificial {I}ntelligence: {P}ublic consultation
  towards a {E}uropean approach for excellence and trusts.
\newblock
  http://ec.europa.eu/digital-single-market/en/news/white-paper-artificial-intelligence-public-consultation-towards-european-approach-excellence.
\newblock [Online; accessed May 24, 2021].

\bibitem[{European Commission}, 2021]{eu2021proposal}
{European Commission} (2021).
\newblock {P}roposal for a {R}egulation of the {E}uropean {P}arliament and the
  {C}ouncil: {L}aying {D}own {H}armonised {R}ules on {A}rtificial
  {I}ntelligence ({A}rtificial {I}ntelligence {A}ct) and {A}mending {C}ertain
  {U}nion {L}egislative {A}cts.
\newblock
  http://digital-strategy.ec.europa.eu/en/library/proposal-regulation-laying-down-harmonised-rules-artificial-intelligence-artificial-intelligence.
\newblock [Online; accessed May 17, 2021].

\bibitem[Evans, 2020]{cnbc2018facebook}
Evans, D. (2020).
\newblock Facebook is still struggling with election manipulation.
\newblock
  http://www.cnbc.com/2020/09/19/2020-presidential-election-facebook-and-information-manipulation.html.
\newblock [Online; accessed May 17, 2021].

\bibitem[Glauner, 2021a]{glauner2021artificial}
Glauner, P. (2021a).
\newblock {A}rtificial {I}ntelligence in {H}ealthcare: {F}oundations,
  {O}pportunities and {C}hallenges.
\newblock In {\em Digitalization in Healthcare: Implementing Innovation and
  Artificial Intelligence}, pages 1--15. Springer.

\bibitem[Glauner, 2021b]{glauner2021stellungnahme}
Glauner, P. (2021b).
\newblock {S}chriftliche {S}tellungnahme f{\"u}r das am 06.05.2021
  stattfindende gemeinsame {F}achgespr{\"a}ch der {A}ussch{\"u}sse für die
  {A}ngelegenheiten der {E}urop{\"a}ischen {U}nion des {D}eutschen
  {B}undestages und der franz{\"o}sischen {A}ssembl{\'e}e nationale zur
  {P}olitik der {EU} im {B}ereich der {K}ünstlichen {I}ntelligenz ({KI}) und
  insbesondere dem {V}erordnungsvorschlag der {E}urop{\"a}ischen {K}ommission
  zu {KI} ({COM}(2021) 206 final).
\newblock http://www.glauner.info/expert-evidence.
\newblock [Online; accessed May 17, 2021].

\bibitem[Heikkil{\"a}, 2021]{politico2021AI}
Heikkil{\"a}, M. (2021).
\newblock 6 key battles ahead for {E}urope’s {AI} law.
\newblock
  http://www.politico.eu/article/6-key-battles-europes-ai-law-artificial-intelligence-act/.
\newblock [Online; accessed May 17, 2021].

\bibitem[Herwartz, 2021]{handelsblatt2021Neu}
Herwartz, C. (2021).
\newblock {N}eue {EU}-{R}egeln für {KI} k{\"o}nnten f{\"u}r {E}uropa zum
  {N}achteil werden – {S}orge über ausbleibende {I}nvestitionen.
\newblock
  http://amp2.handelsblatt.com/politik/international/kuenstliche-intelligenz-neue-eu-regeln-fuer-ki-koennten-fuer-europa-zum-nachteil-werden-sorge-ueber-ausbleibende-investitionen/27113210.html.
\newblock [Online; accessed May 17, 2021].

\bibitem[Kelly, 2021]{sciencebusiness2021call}
Kelly, {\'E}. (2021).
\newblock {C}all for a `{CERN}' for {AI} as {P}arliament hears warnings on risk
  of killing the sector with over-regulation.
\newblock
  http://sciencebusiness.net/news/call-cern-ai-parliament-hears-warnings-risk-killing-sector-over-regulation.
\newblock [Online; accessed May 17, 2021].

\bibitem[{KI Bundesverband}, 2021]{kib2021position}
{KI Bundesverband} (2021).
\newblock {P}osition {P}aper on {EU}-{R}egulation of {A}rtificial
  {I}ntelligence by the {G}erman {AI} {A}ssociation.
\newblock
  http://ki-verband.de/wp-content/uploads/2021/03/Final\_Regulierung\_compressed-1-1.pdf.
\newblock [Online; accessed May 17, 2021].

\bibitem[Lee, 2018]{lee2018ai}
Lee, K.-F. (2018).
\newblock {\em {AI} {S}uperpowers: {C}hina, {S}ilicon {V}alley, and the {N}ew
  {W}orld {O}rder}.
\newblock Houghton Mifflin Harcourt.

\end{thebibliography}

\end{document}